# Looking for New Physics in b-decays with LHCb

F. Teubert
*CERN, PH Department, Geneva 1211, Switzerland*
*On behalf of the LHCb Collaboration.*

This article is a short and non-exhaustive summary of the prospects to find New Physics with LHCb as was presented at the HCP conference at Toronto on August 26$^{th}$ 2010.

## 1. INTRODUCTION

LHCb is a dedicated *b* and *c*-physics precision experiment at the LHC that will search for New Physics (NP) beyond the Standard Model (SM) through the study of very rare decays of charm and beauty-flavoured hadrons and precision measurements of CP-violating observables. At present, one of the most mysterious facts in particle physics is that, on the one hand, NP is expected in the TeV energy range to solve the hierarchy problem, but, on the other hand, no signal of NP has been detected through precision tests of the electroweak theory at LEP, SLC, Tevatron or through flavour-changing and/or CP-violating processes in *K* and *B* decays. In the last decade, experiments at *B* factories have confirmed the validity of the SM within the accuracy of the measurements. The domain of precision experiments in flavour physics has been extended from the kaon sector to the richer and better computable realm of *B* decays. The main conclusion of the first generation of *B*-decay experiments can be expressed by saying that the Cabibbo-Kobayashi-Maskawa (CKM) description [1][2] of flavour-changing processes has been confirmed in $b \rightarrow d$ transitions at the level of 10-20% accuracy, (see Figure 1 left). However, NP effects can still be large in $b \rightarrow s$ transitions (see Figure 1 right), modifying the $B_s$ mixing phase $\beta_s$ measured from $B_s \rightarrow J/\psi\phi$ decays [3], or in channels dominated by other loop diagrams, like, for example, the very rare decay $B_s \rightarrow \mu^+\mu^-$, e.g. via Higgs penguin diagrams [4][5].

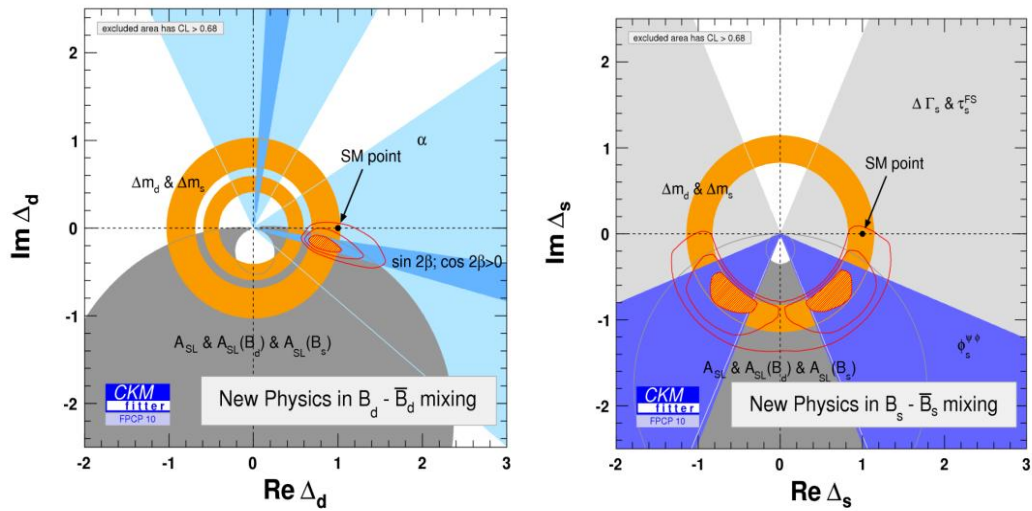

Figure 1: Assuming the effect of NP can be parameterized as a multiplicative factor (Δ) to the SM mixing amplitude ($M_{12}$), this plot shows the imaginary and real parts of this factor allowed by the current measurements, as obtained by the CKMfitter collaboration [6]. The $B_d$ mixing is shown on the left plot and on the right the $B_s$ mixing.

In fact, the recent measurements of large values of $\beta_s$ from $B_s \rightarrow J/\psi(\mu\mu)\phi$ at CDF/D0 [7] and the measurements of an anomalous inclusive dimuon charge asymmetry at D0 [7] may point to large NP contribution to the $B_s$ mixing, as the probability of all these measurements to be consistent with the SM is only few percent. From Figure 1 (left) one can also see that the compatibility of the $B_d$ mixing observables is also only few percent, however in this case the effects of NP are constrained to be much smaller due to the precision of the measurements.

With the startup of LHC and the impressive performance already achieved by LHCb, see [8] and [9], the prospects of clarifying these hints from the Tevatron, or finding clear evidence for NP in charm and/or beauty hadron decays in 2011 are very exciting, indeed.

## 2. LHCb PERFORMANCE

When more than 20 years ago some pioneers thought to have a specialized B-physics detector at the LHC, which eventually became LHCb, the main reason was the enormous $b\bar{b}$ cross section expected. So far, in all the MC studies performed by LHCb the assumption was: 500 $\mu$b (at $\sqrt{s}$ =14 TeV) or 220 $\mu$b (at $\sqrt{s}$ =7 TeV). However, the theoretical uncertainties in these predictions were very large. It certainly was very good news when the first measurements at LHCb of this cross section using semileptonic B decays and displaced J/$\psi$ leptonic decays provided the measurement 298±15±43 $\mu$b, see [8][9], which is slightly larger than assumed in the MC sensitivity studies. At the time of writing these proceedings LHC has already delivered more than 15 pb$^{-1}$ of proton-proton collisions at $\sqrt{s}$ =7 TeV.

Another critical aspect to achieve the LHCb Physics program is the performance of the trigger system. The first triggering level (L0) is a hardware trigger using the Calorimeters, Muon chambers and few specialized VELO sensors. It has been validated early on with the initial data taking and it performs close to MC expectations. The second triggering level (HLT) is a software trigger using all detector information. The flexibility of the HLT system has allowed adapting it to the pileup conditions that LHC is running, and the performance has been validated using high statistics samples of J/$\psi$ leptonic decays and D hadronic decays. Amazingly the performance is very close to MC expectations. Hence two of the more critical aspects of the LHCb Physics program, i.e. the $b\bar{b}$ cross section and the trigger performance, turn out to be close (or even better) than MC expectations.

The offline reconstruction of the triggered events is also performing amazingly well, with tracking efficiencies and lepton/photon identification very close to MC expectations. The momentum and vertex resolution achieved by LHCb is by far the best achieved at the LHC. For instance two body B decays have been measured with a ~35 MeV mass resolution, or the impact parameter for very high $P_T$ tracks has been determined with a precision of ~16 $\mu$m. However, residual misalignments are still limiting the mass and impact parameter resolutions to ~30-50% worse than expected. At the time of writing these proceedings, the enormous effort of the LHCb alignment team has already improved the performance quoted at this conference, and the resolution is now within ~10% of the MC expectations.

Although there are still plenty of work ahead us, for instance the tagging performance is just starting to be commissioned, as far as we can see the sensitivities quoted using MC simulations do not seem to be far from reality.

## 3. EXAMPLES OF INDIRECT SIGNALS OF NEW PHYSICS

There are many possibilities where LHCb can show indirect signs of NP. In the following, I just give few examples where LHCb has a large potential to unveil the existence of NP.

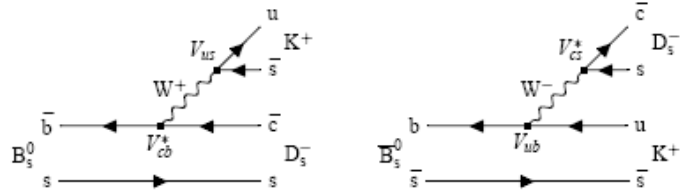

Figure 2: Feynman diagrams for $B_s \to D_s^- K^+$ and $\overline{B}_s \to D_s^- K^+$ in the SM.

### 3.1. New CP phases in B mixing: γ at tree level vs β from loops.

The phase of the $V_{td}$ coupling (-β), which in the SM determines the amount of CP violation measured in the interference of CP eigenstates B decays, has been measured very precisely by the B factories, notably using the decay B→J/ψ$K_s$. However, if NP is present in B mixing (which is not excluded at the 10-20% level in Figure 1), the measurements at the B factories are really a measurement of sin(2β+φ$^{NP}$). If we assume that at the energy scale of B-physics, NP enters only through loop corrections, we can disentangle the effect of NP by measuring sin(2β) using only processes dominated by SM tree diagrams. As there is not really such a process, the alternative is to measure the angle γ (the phase of $V_{ub}$ in the SM) with tree dominated processes. The comparison of the indirect determination of γ using loop dominated processes (68±4)° with the tree dominated one may reveal the existence of NP phase in the B mixing system.

The weak phase γ+β$_s$ can be measured at LHCb from a time-dependent CP asymmetry analysis of the decay $B_s \to D_s^\mp K^\pm$ (see Figure 2)[10]. As a $B_s$ and a $\overline{B}_s$ can both decay as $D_s^+K$, there is interference between the $B_s$ decays where it has or not oscillated. The phase in the $B_s$ mixing can be measured at LHCb using other processes (see next section). The main issue is separating the decay $B_s \to D_s^\mp K^\pm$ from the very similar decay $B_s \to D_s^\mp \pi^\pm$, with a ~15 times larger branching ratio. Here, use of the RICH system crucially allows the background contamination to be reduced. The unique capability of the LHCb trigger to select hadronic decays is instrumental to collecting sufficient statistics. Extrapolating the observed yield so far, with 1 fb$^{-1}$ by the end of 2011, LHCb should have collected ~3.5k signal events with an estimated precision on γ of ~14°.

Alternatively, γ can be extracted from $B^\pm$ and B decays to open charm, as done by the B factories, where a precision of ~20° was achieved through this method. From the experimental point of view, this analysis is much simpler as no flavor tagging or time-dependent analysis is required. Various methods using $B^\pm \to (D/\overline{D})K^\pm$ and B→$\overline{D}$ K* and B→$D_{CP}$K* decays have been proposed [11][12][13][14][15][16]. Extrapolating the observed yields so far, with 1 fb$^{-1}$ by the end of 2011, LHCb should have collected ~25 B→$D_{CP}(\pi\pi)K^*(K\pi)$ as the lowest yield and ~200k B⁻→D(Kπ)K⁻ in the most favored mode. That statistics should be enough to determine $\gamma$ with a precision of ~8°. This precision should allow an interesting comparison with the indirect determination (68±4)° and may reveal hints of NP at the ~20% level in the B mixing system.

### 3.2. New CP phases in $B_s$ mixing: $\beta_s$ and $a_{sl}$

Indeed looking for new phases in B mixing requires high precision given the current constraints from B-factories; however the situation is very different in the $B_s$ mixing induced CP asymmetries. First measurements from CDF/D0 of this phase using the time dependent analysis of the decay $B_s$→J/ψϕ shows tantalizing hints, see reference [7]. Moreover, the recent D0 measurement of an anomalous dimuon charge asymmetry points in the same direction [7]. The probability that the SM is consistent with all these observations is only few percent.

The J/ψϕ final state is a sum of CP eigenstates and each contribution can be disentangled on a statistical basis. This is realized by performing an analysis based on the so-called transversity angle, defined as the angle between the positive lepton and the ϕ decay plane in the J/ψ rest frame. The phase $\beta_s$ is determined through a simultaneous fit to the proper time and transversity angle distributions, as well as to the proper time distribution of a control sample of $B_s$→$D_s\pi$ decays, which is used to extract $\Delta m_s$ and the tagging efficiency. Extrapolating the observed yields so far, LHCb expects approximately 2.5k $B_s$→ J/ψϕ signal events in 50 pb$^{-1}$ of data with a background over signal ratio B/S~0.2. The expected statistical sensitivity on $\beta_s$ is ~0.5 radians with these data, if the tagging performance and proper time resolution are not far from expectations, which is similar to CDF's precision using 5.2 fb$^{-1}$ of data. In Figure 3(left) one can see how a measurement from LHCb with the data collected in 2010 (~50 pb$^{-1}$) could reinforce (or dismiss) the hints observed by CDF/D0. Certainly, with the data collected in 2011 (~1 fb$^{-1}$) a clear observation will be possible if the phase is as large as indicated by CDF/D0.

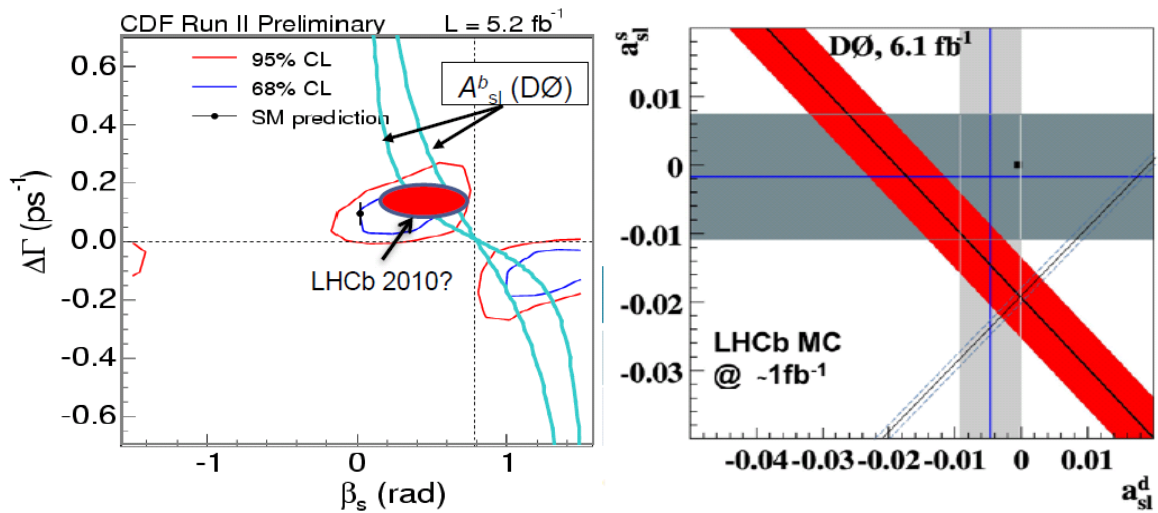

Figure 3: On the left is shown the favored regions in the plane $\beta_s$ vs $\Delta\Gamma$ by the CDF measurements using $B_s$→J/ψϕ and by the D0 measurements of the dimuon asymmetry. A hypothetical measurement of LHCb using 50 pb$^{-1}$ of data in 2010 is also shown. On the right is shown the favored regions in the plane $a_{sl}^d$ vs $a_{sl}^s$ by B-factories and D0 measurements. A hypothetical measurement of LHCb using 1 fb$^{-1}$ of data in 2011 is also shown.

The current measurement by D0 of the inclusive charge dimuon asymmetry [7], deviates from expectations by ~3$\sigma$ assuming the observed dimuons are all originated from b-decays. In this hypothesis, the measurement can be interpreted as the sum of $0494 a_{sl}^s + 0.506 a_{sl}^d$, which corresponds to the red band shown in Figure 3 (right). If the measurement of $a_{sl}^d$ from the B-factories (shown in Figure 3(right) as a vertical band, in good agreement with the SM) is plugged in, the D0 measurement can be interpreted as a ~2$\sigma$ hint for non SM contributions in the $B_s$ mixing: $a_{sl}^s = -(1.46\pm0.75)\times10^{-2}$.

The LHC is a proton-proton collider which implies a difficult to predict production asymmetry of order ~$10^{-2}$. Moreover, as in the case of D0, LHCb expects some detector asymmetries at this level that are controlled by regularly switching the magnet polarity. These facts make the inclusive approach at LHCb very difficult. However it looks very promising to pursue an exclusive approach, either using hadronic decays ($B_s \rightarrow D_s\pi$) or semileptonic decays and leave the production asymmetry as an extra free parameter in the fit. In the case of semileptonic decays, in order to control the remaining detector asymmetries it is better to subtract $B_s$ and B decays, hence measuring a linear combination of $a_{sl}^s - a_{sl}^d$ which is orthogonal to the D0 measurement, see Figure 3 (right). LHCb expects to measure this linear combination using semileptonic decays with a precision of ~$3\times10^{-3}$ already with the first 50 pb$^{-1}$ of data that we may collect in 2010.

### 3.3. New Lorentz structure: B→K*µ$^+$µ$^-$

In the SM, the decays $b \rightarrow sl^+l^-$ cannot occur at tree level but only through electroweak penguin diagrams with small branching fractions, e.g. BR($B \rightarrow K^*\mu\mu$)~$1.2 \times 10^{-6}$. The $B \rightarrow K^*\mu\mu$ channel is well suited to searches for NP, because most NP scenarios make definite predictions for the forward-backward asymmetry $A_{FB}$ of the angular distribution of the $\mu$ relative to the $B$ direction in the di-muon rest frame as a function of the di-muon invariant mass $m_{\mu\mu}$. In particular, the value of $m_{\mu\mu}$ for which $A_{FB}$ becomes zero is predicted with small theoretical uncertainties and may thus provide a stringent test of the SM [17]. The $B$ factories and CDF have succeeded in measuring $A_{FB}$ vs di-lepton $q^2$ as shown in Figure 4. The measurements have a tendency to favour non SM values for some combination of the Wilson coefficients, notably the coefficient related to the magnetic operator ($O_7$). LHCb expects to have a similar precision with ~150 pb$^{-1}$ of data and certainly, with 1 fb$^{-1}$ expected in 2011, should be able to see a clear sign for NP if the effect is as large as may be hinted at by the measurements today (see Figure 4).

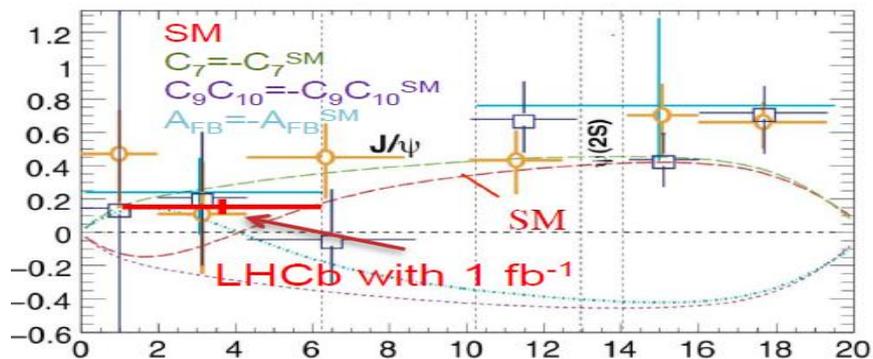

Figure 4: $A_{FB}$ vs dimuon $q^2$ as measured by Belle (orange cercles), BaBar (blue crosses) and CDF (dark blue squares), compared with an hypothetical measurement by LHCb with 1 fb$^{-1}$. The different dashed and dotted lines correspond to predictions with different values of the Wilson coefficients. The red dashed line corresponds to the SM.

## 3.4. New (pseudo-)scalar contributions: $B_s \to \mu^+\mu^-$

The decay $B_s \to \mu^+\mu^-$ has been identified as a very interesting potential constraint on the parameter space of models for physics beyond the SM [4] [5]. The upper limit to the $B_s \to \mu^+\mu^-$ branching ratio measured at CDF [7] is $3.6 \times 10^{-8}$ @90% C.L. The SM prediction is computed [18] to be BR($B_s \to \mu^+\mu^-$) = $(3.2 \pm 0.2)\, 10^{-9}$ using the latest measurement of the $B_s$ oscillation frequency at the Tevatron, which significantly reduces the uncertainties in the SM prediction. Within the SM, this decay is dominated by a "Z/Higgs-penguin" diagram, while the contribution from the "box" diagram is suppressed by a factor $\sim (M_w/m_t)^2$. Moreover, the axial amplitude, which dominates in the SM, is helicity suppressed by a factor $\sim (m_l/m_{Bs})^2$, hence it is very sensitive to any NP with new scalar or pseudoscalar interactions, in particular to any model with an extended Higgs sector. In the MSSM this branching ratio is known to increase as the sixth power of $\tan\beta = v_u/v_d$, the ratio of the two Higgs vacuum expectation values. Any improvement to this limit is therefore particularly important for models with large $\tan\beta$.

The large background expected in the search for the decay $B_s \to \mu^+\mu^-$ is largely dominated by random combinations of two muons originating from $b$ decays. This background can be kept under control by exploiting the excellent tracking and vertexing capabilities of LHCb. Specific decays, such as $B_{(s)} \to h^+h^-$, where the hadrons ($h$) are misidentified as $\mu$, or $B_c \to J/\psi\, (\mu^+\mu^-)\, \mu\nu$, do not contribute to the background at a significant level, compared to the combinatorial background, due to the very low $\mu$ misidentification rate and excellent invariant mass resolution. The analysis in LHCb does not depend critically on the quality of the MC simulation, as the background probability can be computed from events in the mass sidebands, and the signal probability can be computed thanks to the unique possibility at LHCb to use large samples of $B_{(s)} \to h^+h^-$ decays. Precisely the clean observation of these decays with the initial data, and the properties measured being very close to expectations gives us confidence in the LHCb potential for this search. The events are classified according to the invariant mass and the Geometrical Likelihood (GL), which is just a multivariate analysis that mainly takes into account quantities related to the vertex detector. In Figure 5 (left) one can see the distribution of the data in these two variables. While the background is expected to accumulate at low values of GL and have a linear distribution in mass, the signal should be uniform in GL and peak around the $B_s$ mass. The initial data confirms the expected distributions from MC. Figure 5 (right) shows the extrapolation of the expected limit on BR($B_s \to \mu^+\mu^-$) as a function of the luminosity. It is clear than LHCb should enter "terra incognita" with very little integrated luminosity.

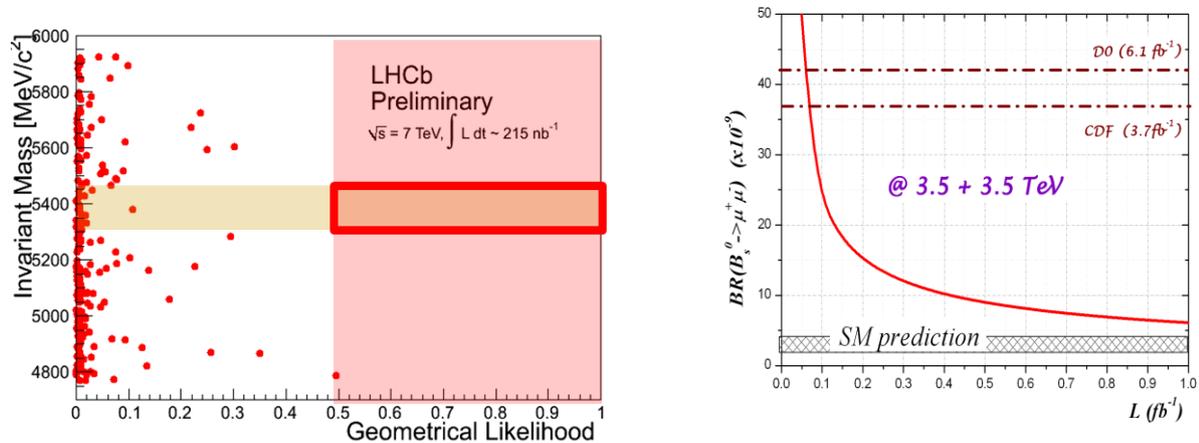

Figure 5: On the left, Data distribution on the GL vs invariant mass plane using the first 0.2 pb$^{-1}$. On the right the expected 90% C.L. limit as a function of integrated luminosity compared with the limits obtained at the Tevatron.

## 4. OUTLOOK AND CONCLUSIONS

The LHCb experiment will certainly contribute significantly to the overall LHC effort to find and study physics beyond the SM. In fact, the B-physics program is not much affected by LHC running at half the nominal energy. A few highly sensitive b→s observables are accessible in 2010/11: the decay $B_s \to \mu^+\mu^-$ may indicate the existence of non-SM higgses, the decay B→K*μμ may indicate modifications to the SM couplings structure and new phases appearing in the $B_s$ mixing may be seen using the decay $B_s$→J/ψϕ and measuring $a_{sl}$ using semileptonic decays. In a somewhat longer time scale the existence of NP phases contributing to the B mixing may be revealed by measuring γ from tree decays.

The LHCb detector is performing already very close to expectations. Stay tuned for what LHCb has to say on the existence of NP in the forthcoming months.

## References

[1] N.Cabibbo, *Phys. Rev. Lett.* **10,** 531 (1963).

[2] M.Kobayashi and T.Maskawa, *Prog. Theor. Phys.* **49,** 652-657 (1973).

[3] A.S.Dighe, I.Dunietz, H.J.Lipkin and J.L.Rosner, *Phys. Lett. B* **369,** 144-150 (1996).

[4] A. Dedes, H.K.Dreiner and U.Nierste, *Phys. Rev. Lett.* **87,** 251804 (2001).

[5] C.Huang, L.Wei, Q.Yan and S.Zhu, *Phys. Rev.* **D63,** 114021 (2001).

[6] J. Charles *et al.* (CKMfitter Collaboration), *Eur. Phys. J.* **C41,** 1, hep-ph/0406184 (2005).

[7] B. Lee, Contribution to these proceedings.

[8] G. Carboni, Contribution to these proceedings.

[9] B. Viaud, Contribution to these proceedings.

[10] R.Aleksan, I.Dunietz and B.Kayser, *Z. Phys.* **C54,** 653 (1992).

[11] D.Atwood, I.Dunietz and A.Soni, *Phys. Rev. Lett* **78***,* 3257 (1997).

[12] M.Gronau and D.London, *Phys. Lett. B* **253**, 483 (1991).

[13] M. Gronau and D.Wyler, *Phys. Lett. B* **265**, 172 (1991).

[14] M.Gronau, *Phys. Rev.* **D58**, 037301 (1998).

[15] D.Atwood, I.Dunietz and A.Soni, *Phys. Rev.* **D63**, 036005 (2001).

[16] A.Giri, Y.Grossman, A.Soffer and J.Zupan, *Phys. Rev. D* **68**, 054018 (2003).

[17] A.Ali, P.Ball, L.T.Handoko and G.Hiller, *Phys. Rev. D* **61**, 074024 (2000).

[18] A.J.Buras, M.V.Carlucci, S.Gori and G.Isidori, *arXiv:1005.5310*